\documentclass[%
reprint,
superscriptaddress,
amsmath,amssymb,
aps,
prb,
showkeys,               
]{revtex4-2}

\usepackage{graphicx}
\usepackage{bm}

\usepackage{siunitx}
\usepackage{chemformula}
\usepackage{enumitem}
\usepackage{xcolor}

\begin{document}


\title{Probing magnetic properties at the nanoscale: In-situ Hall measurements in a TEM} 

\author{Darius Pohl}
 \email{darius.pohl@tu-dresden.de}
 \affiliation{Dresden Center for Nanoanalysis (DCN), cfaed, TU Dresden, 01062 Dresden, Germany}
\author{Yejin Lee}
 \affiliation{Leibniz Institute for Solid State and Materials Research, IFW Dresden, 01062 Dresden, Germany}
\author{Dominik Kriegner}
\affiliation{Institute of Physics, Academy of Sciences of the Czech Republic, Cukrovarnick\'a 10, 162 00 Praha 6, Czech Republic}
 \affiliation{Institute of Solid State and Materials Physics, TU Dresden, 01062 Dresden, Germany}
\author{Sebastian Beckert}
 \affiliation{Institute of Solid State and Materials Physics, TU Dresden, 01062 Dresden, Germany}
\author{Sebastian Schneider}
 \affiliation{Dresden Center for Nanoanalysis (DCN), cfaed, TU Dresden, 01062 Dresden, Germany}
\author{Bernd Rellinghaus}
 \email{bernd.rellinghaus@tu-dresden.de}
 \affiliation{Dresden Center for Nanoanalysis (DCN), cfaed, TU Dresden, 01062 Dresden, Germany}
\author{Andy Thomas}
 \email{andy.thomas@tu-dresden.de}
 \affiliation{Institute of Solid State and Materials Physics, TU Dresden, 01062 Dresden, Germany}
 \affiliation{Leibniz Institute for Solid State and Materials Research, IFW Dresden, 01062 Dresden, Germany}

\date{\today}

\begin{abstract}
We report on advanced in-situ magneto-transport measurements in a transmission electron microscope. The approach allows for concurrent magnetic imaging and high resolution structural and chemical characterization of the same sample. Proof-of-principle in-situ Hall measurements on presumably undemanding nickel thin films supported by micromagnetic simulations reveal that in samples with non-trivial structures and/or compositions, detailed knowledge of the latter is indispensable for a thorough understanding and reliable interpretation of the magneto-transport data. The proposed in-situ approach is thus expected to contribute to a better understanding of the Hall signatures in more complex magnetic textures.
\end{abstract}

\keywords{In-situ TEM, Hall effect, Lorentz TEM, Micromagnetic simulations}
\maketitle

\subsection{Introduction}
The field of spinelectronics (or spintronics for short) relates magnetic phenomena to transport properties. Examples are the giant magnetoresistance \cite{baibich_giant_1988,binasch_enhanced_1989}, the tunneling magnetoresistance \cite{julliere_tunneling_1975,moodera_large_1995,miyazaki_giant_1995} or the recently reported topological Hall effect \cite{Neubauer:2009,Bruno:2004}. In almost all of these investigations, additional support from magnetometric measurements or magnetic imaging is needed to explain the observed results. However, since the stability of magnetic textures depends on the sample geometry, a correlation of magnetotransport data with, e.g., transmission electron microscopy (TEM) images or the aforementioned magnetometry data is problematic if not obtained from {\em identical} samples. 

We have therefore developed an in-situ measurement platform that bridges this gap and allows for in-situ magnetotransport measurements in a TEM. The approach enables (i) measurements of the Hall effect and magnetoresistance, (ii) structural and chemical characterization using TEM, electron energy loss spectroscopy (EELS) and energy dispersive X-ray spectroscopy (EDS), as well as (iii) magnetic imaging by Lorentz-TEM (LTEM) on the same sample. Such combined experiments have already proven very valuable in unraveling the structure-property relations, e.g, in nanomagnets \cite{Schneider:2016,Wicht:2013} and will pave the way to a better understanding of complex magnetic textures, e.g, in skyrmionic materials \cite{skyrme_thr_non-linear_1961,kiselev_chiral_2011}. 

First, we present our in-situ TEM setup, which allows for automated field sweeps, the collection of longitudinal and transversal voltages while applying an electric current, and simultaneous magnetic imaging using LTEM. TEM investigations of electrically biased samples are facilitated by modern in-situ holders that are equipped with electrical feedthroughs into the high vacuum of the microscope column \cite{Almeida:2020,Romero:2019,Wittig:2017} thereby opening the possibility for a variety of in-situ electrical characterizations. Recent work in this direction was limited to two-terminal applications, e.g., for the current induced motion of domain walls or skyrmions \cite{Junginger:2007,Yu:2017} or the measurement of the linear magneto-resistance \cite{Tang:2023}. Here, we extend these approaches to multi-terminal measurements and focus on the combined measurement of the Hall effect and magnetic imaging using LTEM. 

The Hall effect in solids, first discovered by Edwin H.\ Hall in 1879 \cite{Hall:1879}, describes the generation of a voltage perpendicular to both the applied current and magnetic field. In ferromagnets, the resulting Hall resistance is generally composed of the ordinary Hall effect (OHE) caused by the Lorentz force and the anomalous Hall effect (AHE) resulting from different scattering mechanisms (intrinsic, skew, side jump) in the magnetic material \cite{Nagosa:2010}. Its significance for the determination of both the density and mobility of charge carriers and more recently its relation to topological phenomena renders the Hall resistance a key property in modern solid state physics. 

Lorentz transmission electron microscopy (LTEM) is an advanced imaging technique capable of visualising magnetic structures such as domain walls, skyrmions, or even more complex magnetic textures. The contrast in LTEM images arises from the phase shift imposed on the electron wave by virtue of the in-plane components of the magnetic induction in the sample via the Lorentz force. This weak magnetic phase shift can be converted into visible image contrast by defocusing the sample. Besides the mere visualization of microscopic and nanoscopic ma\mbox{gne}tic features, image reconstruction techniques such as provided by the transport of intensity equation (TIE) even allow for the quantification of the in-plane magnetic induction in the sample \cite{Zweck:2016}. Besides, varying magnetic fields or temperatures can be used as external stimuli \cite{Zhang:2020} to create or annihilate complex magnetic textures in order to study their stability and dynamics. 

Nickel -- a well understood 3d ferromagnet -- has been chosen as a model system to demonstrate the feasibility of our in-situ Hall measurements in a TEM. We will show that even in this presumably simple case, detailed knowledge of the structure and composition of the sample is indispensable to reliably interpret the measured magnetotransport data thereby highlighting the added value of the approach.


\subsection{Sample preparation and magnetotransport measurement setup}
\label{setup}
\begin{figure}[tbp]
\includegraphics[width=7cm]{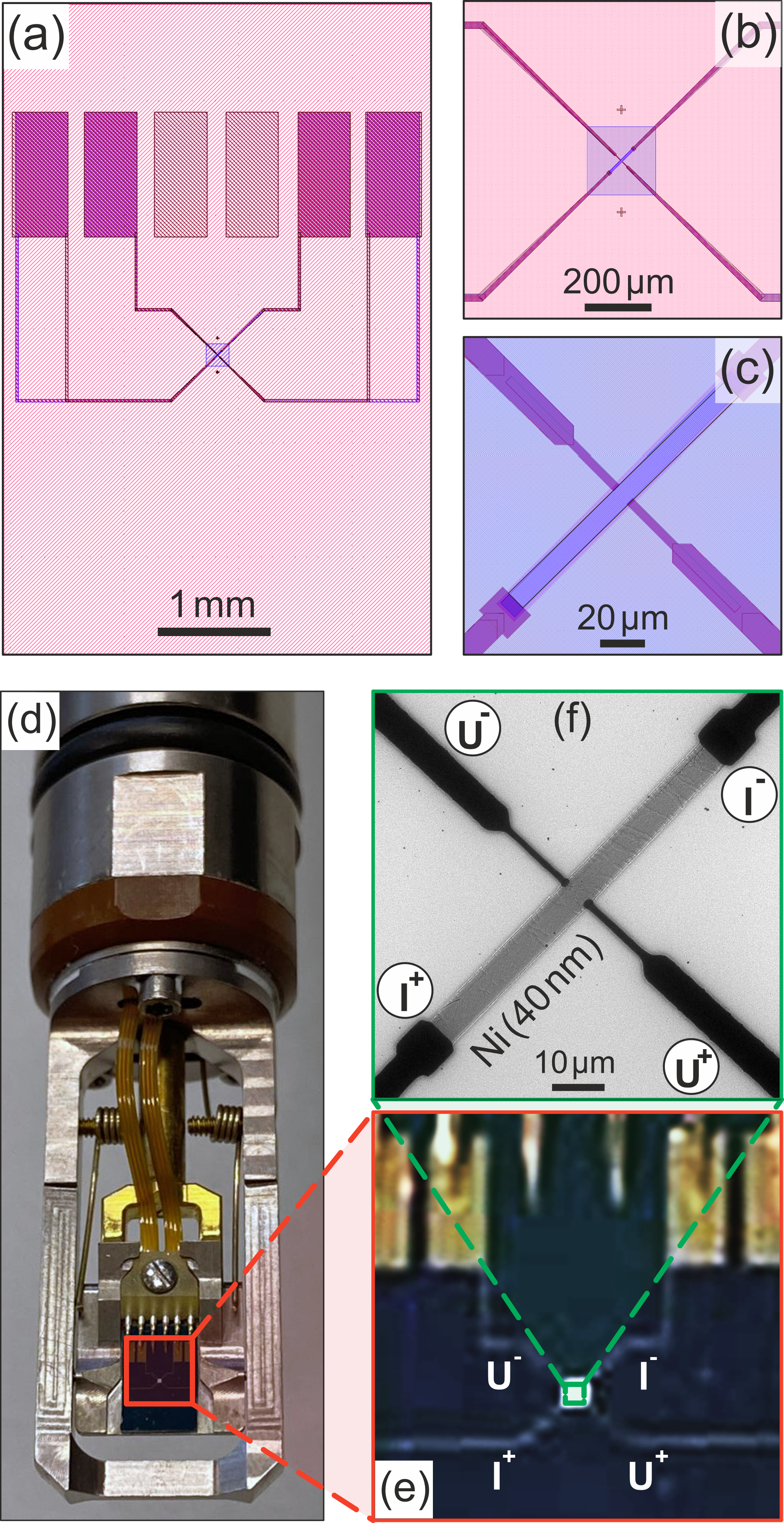}
\caption{Experimental setup for in-situ magneto-transport measurements in the transmission electron microscope (TEM): (a) Layout of the measurement chip (3.75x\SI{5.7}{mm^2}). (b,c) Magnified section of the layout of the same chip with four gold leads readily contacted to the Ni film deposited across an electron transparent \ch{Si3N4} window in its center. (d) Protochips Fusion Select in-situ TEM holder with six electrical feedthroughs, four of which are connected via gold-plated spring contacts to the contact pads on the chip. (e) Magnified image of the contact pads and leads to the Ni sample in the center of the window. (f) TEM image of the Hall bar structure of the Ni film after insertion into the microscope.}
\label{fig:chip_carrier}
\end{figure}

In-situ magnetotransport measurements in a TEM necessitate the use of electron-transparent samples and substrates. We use homemade measurement chips that are adapted to the needs of our Protochips Fusion Select in-situ TEM holder. 
Figs.\ \ref{fig:chip_carrier}(a)-(c) show the computer-aided design of the chips with four rectangular contact pads (dark purple) for the spring contacts of the holder (incl.\ two optional pads, not used here), the electrical leads from the pads to the Ni film (dark purple), an electron-transparent square window in the center (blueish), and finally the bar-like Ni film in the middle of this window. To realize this concept, \ch{Si3N4\slash Si\slash Si3N4} trilayers with layer thicknesses of \SI{100}{nm}, \SI{300}{\micro m} and \SI{100}{nm} are cut to pieces of \SI{3.75}{mm} by \SI{5.7}{mm} in size and subsequently cleaned with acetone and isopropanol. A maskless aligner and conventional optical lithography are then used to define a square window at the front side of the substrate (in the direction of the electron beam) that carries the thin film sample of Ni. From the back side of the trilayer, \ch{Si3N4} is removed by reactive ion etching using \ch{CF4}, and, afterwards, Si is etched with 40\% \ch{KOH} until at the front side a square window of \ch{Si3N4} with an edge length of \SI{200}{\micro m} is laid open. The likewise prepared \ch{Si3N4} window has a thickness of \SI{100}{nm}. 

Now, the structure to be investigated is placed on the \ch{Si3N4} window. Both, thin film samples and FIB cut lamellae \cite{Geishendorf:2019du,moghaddam_observation_2022} are feasible. For the Hall measurements pursued here, we have prepared an approximately \SI{80}{\micro m} by \SI{8}{\micro m} large nickel rectangle with a nominal thickness of \SI{40}{nm} by magnetron sputtering and a subsequent lift-off process utilizing conventional optical lithography. The design is displayed in Fig.\ \ref{fig:chip_carrier}(c) as the blue bar running from the lower left to the top right of the figure with the four darker structures representing the contact leads. 

Fig.\ \ref{fig:chip_carrier}(d) shows a photograph of the in-situ holder with the readily mounted measurement chip contacted to the outermost four of the six gold plated spring contacts of the holder. Here, only four leads are required for the two longitudinal and the two transversal (Hall) contacts. The electrical leads and contact pads on the substrate are prepared from \ch{Cr\slash Au} bilayers by conventional optical lithography and a lift-off process. Fig.\ \ref{fig:chip_carrier}(f) shows a TEM micrograph of the final sample with the nickel film (dark gray) and the longitudinal and transversal contact leads (black) labelled $\rm I^+, I^-$ and $\rm U^+, U^-$,
indicating the polarities of the constant electrical current $I$ and the transversal (Hall) voltage $U$, respectively.

The magneto-transport measurements were conducted in a similar way to our previously reported experiments (see, e.g., Geishendorf et al.\ \cite{Geishendorf:2019du} or Moghaddam et al.\ \cite{moghaddam_observation_2022}). The supply current was set to \SI{5}{\micro A} using a Keithley 2450 source meter. The transverse (Hall) voltage was measured with a Keithley 2182A nanovoltmeter, while the longitudinal voltage was again monitored by the source meter. The integration time of the nanovoltmeter was set to ten times the power line cycle (\SI{1}{PLC} = \SI{20}{ms}). The magnetic field was controlled via the excitation of the objective lens of the microscope and was subsequently increased in \SI{6}{mT} steps. A waiting time of \SI{0.2}{s/mT} was introduced to avoid picking up voltages induced by the magnetic field change. To evaluate the quality of the in-situ magnetotransport measurements in the TEM, we conducted comparative measurements using the identical source- and nanovoltmeters with the same parameters in our conventional cryostat with a variable temperature insert (VTI) at room temperature \cite{Geishendorf:2019du,moghaddam_observation_2022}. For this comparison, the integration time of the nanovoltmeter was deliberately set to a shorter value as this enhances the noise and thus simplifies a quantitative analysis. 

\begin{figure}[tbp]
\includegraphics[width=7cm]{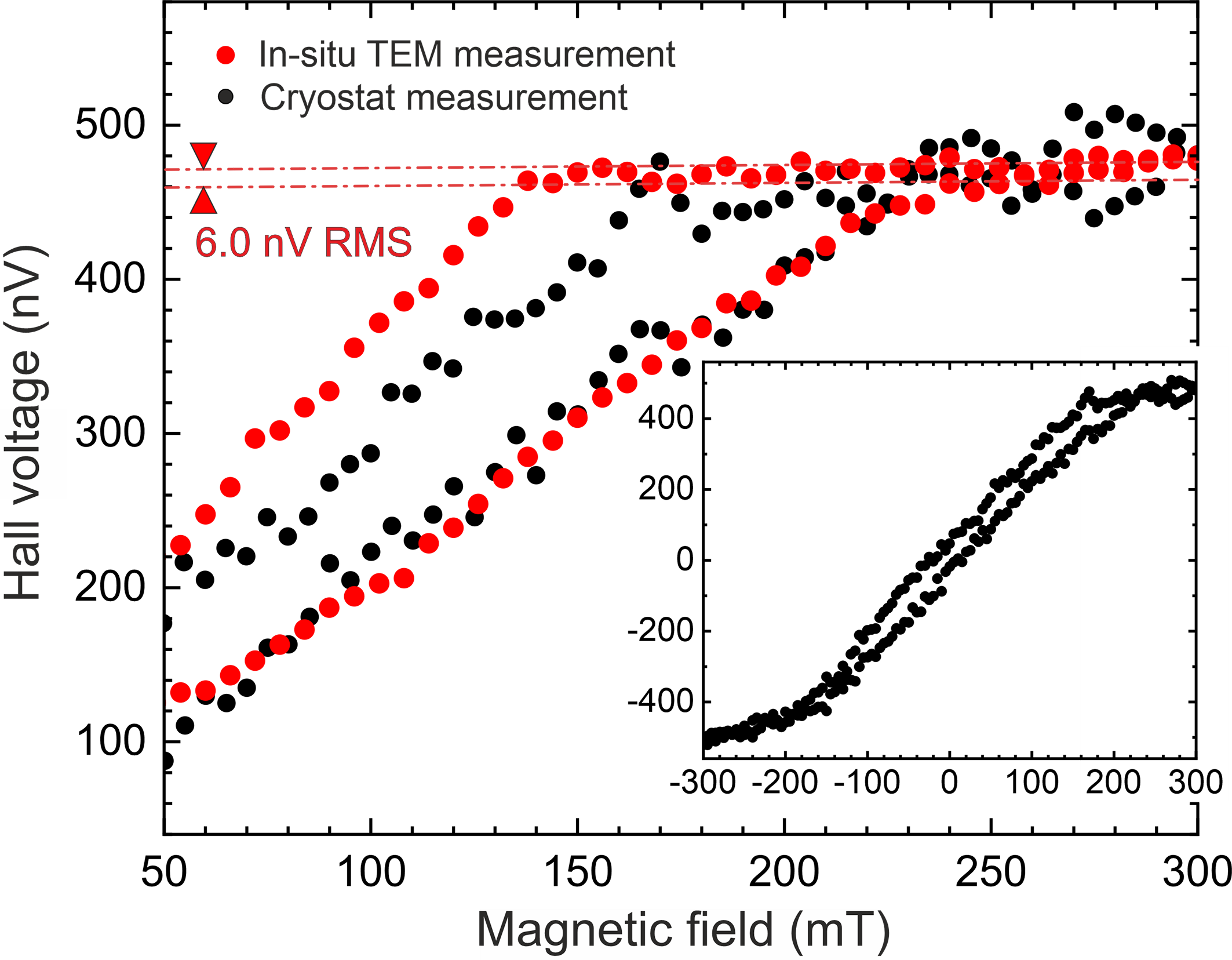}
\caption{Comparison of the noise levels of Hall voltages measured on an identical \SI{40}{nm} thin nickel film (i) in-situ in a TEM (red dots) and (ii) in a conventional cryostat at room temperature (black dots, cf.\ insert for the full hysteresis loop).}
\label{fig:noise}
\end{figure}

The results of these measurement are plotted in Fig.\ \ref{fig:noise}. The root mean square (RMS) values of the noise levels determined in saturation are \SI{6}{nV} and \SI{17}{nV} for the in-situ TEM and cryostat measurements, respectively. Hence, the noise introduced by the in-situ holder with its presumably less well-defined spring contacts turns out to be even lower than that one of our cryostat. The latter is most likely caused by additional thermoelectric voltages due to the large temperature gradient of the sample probe in the cryostat at room temperature. In this sense, the comparison reveals a very good noise figure for our in-situ Hall measurement platform in the TEM. 


\subsection{Microscope control}
Transmission electron microscopy studies have been conducted using our JEOL JEM F200 (cold field emission gun) microscope operated at an acceleration voltage of 200 kV. It is equipped with a GATAN OneView CMOS camera for fast imaging, a GATAN GIF Continuum spectrometer for EELS and a dual \SI{100}{mm^2} windowless silicon drift detector system for EDX. The complete functionality of the microscope and the GATAN OneView camera can be controlled using Python scripting via PyJEM \cite{PyJEM_GIT,PyJEM_JEOL} and the Gatan Microscopy Suite (GMS) \cite{Gatan_GMS}, respectively.

\begin{figure}[tbp]
\includegraphics[width=7cm]{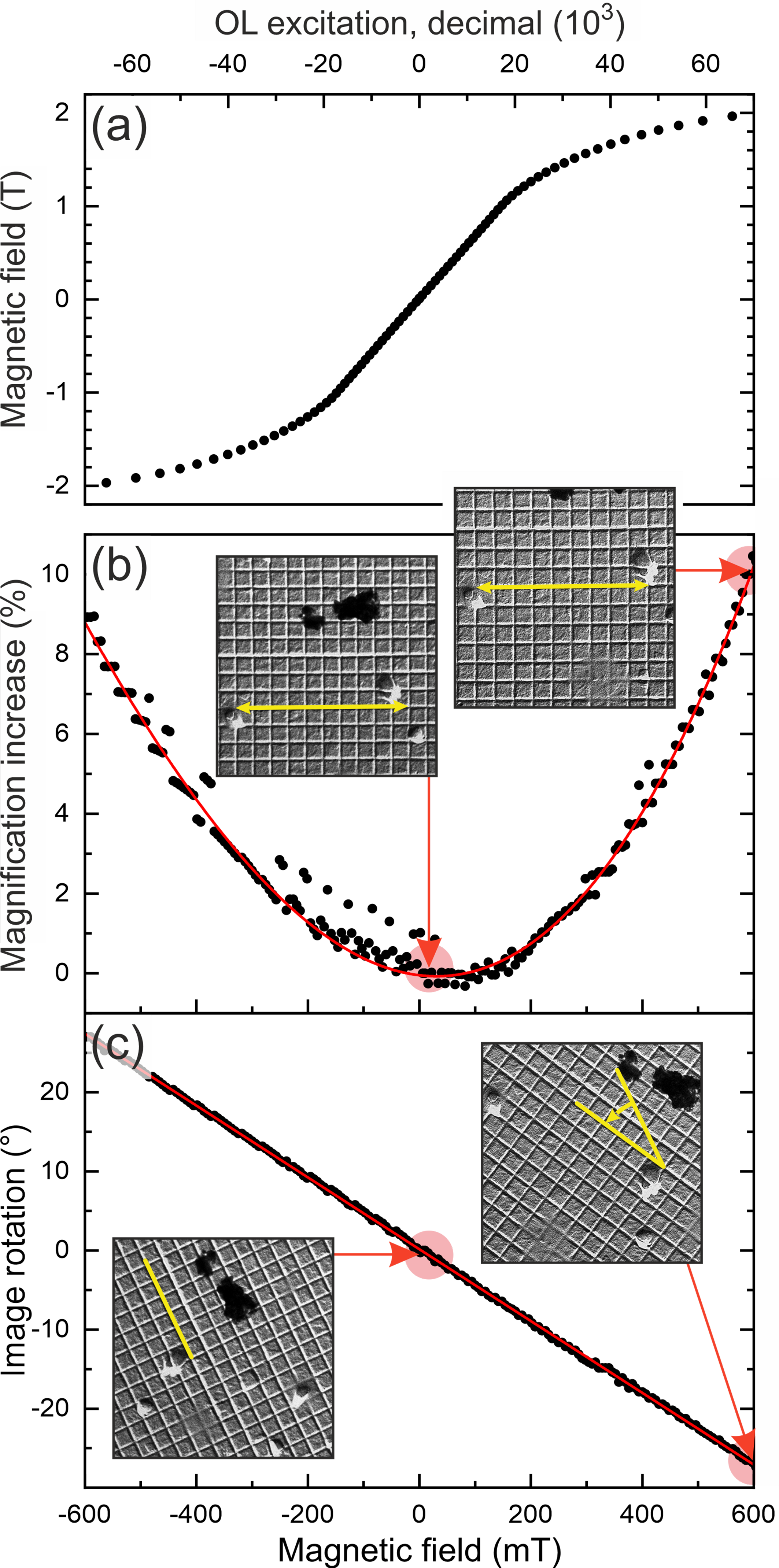}
\caption{Calibration curves for the combined in-situ Hall and LTEM measurements: (a) Magnetic field as function of the objective lens excitation (in decimals as programmed through the microscope's user interface). Negative values represent excitations after manual reversal of the polarity. (b) Relative change of the magnification in the LTEM image (with respect to the value at zero field, black dots) vs.\ applied magnetic field. The red line indicates a polynomial fit to the data utilized for post-processing correction of the images. (c) Magnetic field dependence of the angle of rotation (black dots) of the LTEM images and linear regression of the data (red line). Insert LTEM images of a Au cross-grating sample in (b) and (c) highlight the field effects (at \SI{600}{mT} as compared to \SI{15}{mT}) on the image magnification and rotation, respectively.}
\label{fig:calibration}
\end{figure}

Magnetic imaging via LTEM is done by switching off the objective lens (OL) and refocusing the sample using the first intermediate lens. A magnetic field can be applied perpendicular to the sample plane and controlled by successively increasing the OL excitation. The likewise generated magnetic field at the sample position is measured using a calibrated commercial Hall sensor of the type HE144 from Asensor Technology AB (cf.\ fig.\ \ref{fig:calibration}a). In standard polarity (i.e., as delivered by JEOL), the magnetic field can be varied from \SIrange{10}{1963}{mT}. An external polarity switch for the current supply of the OL is used to manually reverse the direction of the applied field, allowing for magnetic fields also in opposite directions in the range between \SI{-5}{mT} and \SI{-1968}{mT}. 

Besides the controlled application of magnetic fields, the variation of the OL excitation also leads to unwanted beam shifts/tilts, image magnifications/shifts and rotations, and even to focus shifts. In order to compensate for these effects, all of these properties were pre-determined as functions of the OL excitation. The results of these calibrations were then used to implement predictive compensations of beam shift and tilt, image shift, and focus already during the image acquisition. By this automation, LTEM images can be acquired during complete field loops in the range of \SIrange{-600}{600}{mT} without any additional manual corrections. Beyond these field amplitudes, the compensations required during image acquisition would exceed the ranges of the respective correction lenses, which is why the present investigations are restricted to just this range.

Since a predictive compensation of the image magnification and rotation during the acquisition would require complex changes of the complete projective lens system, these latter effects are corrected for via post-processing of the images. In Fig.\ \ref{fig:calibration} we show the measured magnetic field dependence of the magnification (b) and the image rotation (c) together with fits (red lines) to the experimental data (black dots) used for this post-processing. 

The whole measurement is controlled via a self-written, modular Python interface that allows for the flexible integration of modules to control measurement devices such as the source- and nanovolt meters, the JEOL-F200 microscope, or the CMOS camera. E.g., the JEOL-F200 module controls the microscope column by setting the OL excitation and magnetic field and the parameters of the lenses and deflector coils. It also sends trigger signals to the GATAN control PC via Ethernet to initiate the acquisition of LTEM images and provides for metadata such as the detailed microscope settings for inclusion in the image files. This module based scripting allows for a flexible combination of different measurement techniques -- in our case LTEM imaging and magneto-transport measurements -- and can be easily extended for additional functionalities and experiments.

All acquired data (images, data on magnetic fields, voltages and currents, etc.)  can be stored in the standard HDF5 file format \cite{HDF_wiki,HDF_hdfgroup} and are automatically transferred to shared folders on a remote workstation. Once stored, the data is made accessible to all collaborating researchers (via active control of access rights) for further evaluation, publication, and archiving or other research data management purposes.    


\subsection{In-situ experiments with structural analysis and micromagnetic simulations}
For a first proof of concept experiment, Ni films with a thickness of \SI{40}{nm} were prepared as described in section \ref{setup}. 
%
\begin{figure}[tbp]
\includegraphics[width=7cm]{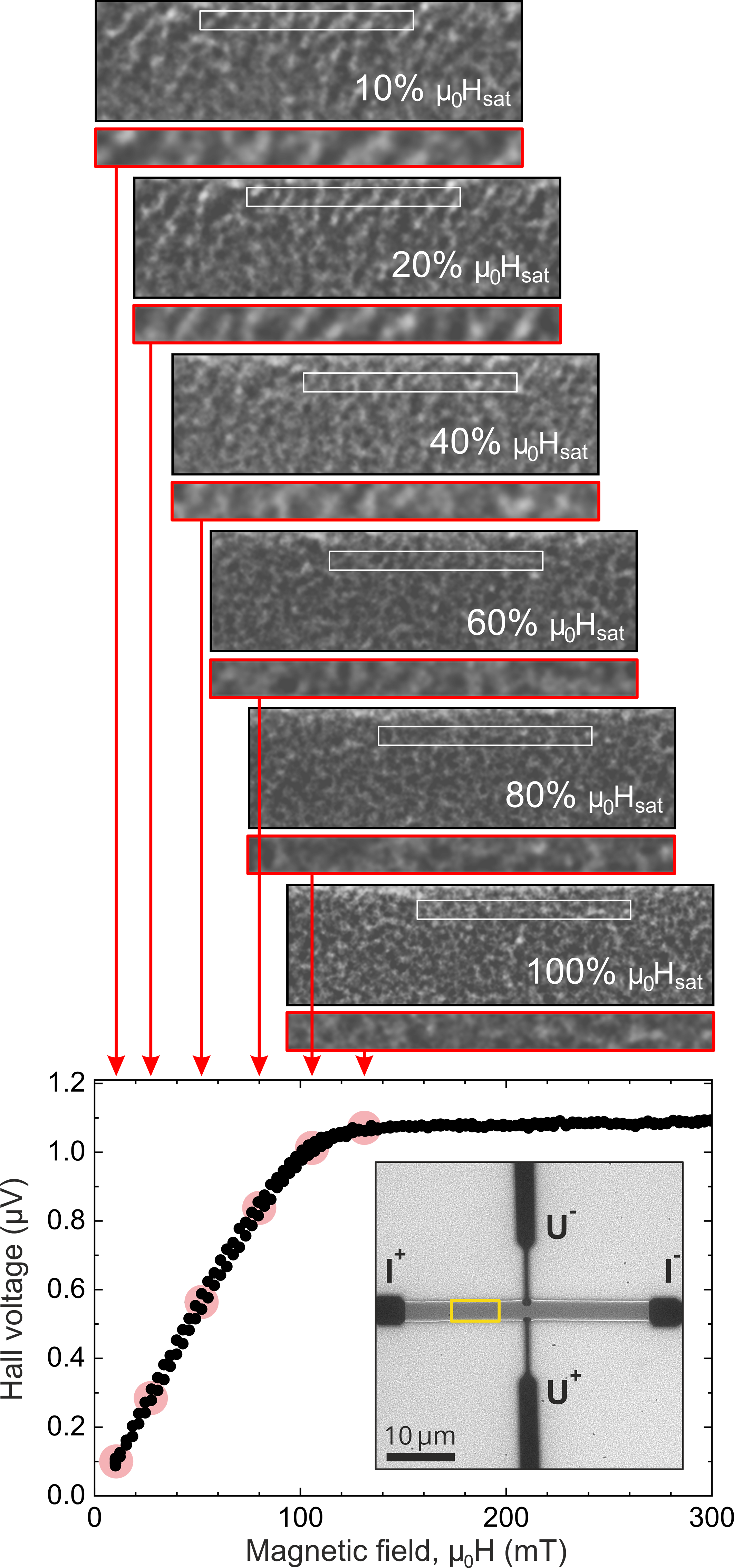}
\caption{Simultaneous LTEM and in-situ Hall measurements: Measured Hall voltage ($U_{H}$) as a function of the applied magnetic field (bottom). The insert shows an overview LTEM image of the nickel Hall bar ($35 \times \SI{3}{\micro m^2}$) with the gold contacts (dark contrast) for current supply and voltage measurements labeled $\rm I^{+/-}$ and $\rm U^{+/-}$, respectively. The yellow rectangle indicates a $\rm \SI{7}{\micro m} \times \SI{2}{\micro m}$ large area, for which sections of the LTEM images are displayed above. Below each of these images, further magnified areas of $\rm \SI{3.5}{\micro m} \times \SI{0.3}{\micro m}$ (framed in red) are displayed for comparison with simulations. For each LTEM image, the corresponding field values are indicated by red arrows to the bottom plot.}
\label{fig:Ni_hall}
\end{figure}
As can be seen from the overview LTEM image in the insert of Fig.\ \ref{fig:Ni_hall}, the current is applied along the long axis of the Ni bar, and the Hall voltage is measured across the bar and thus perpendicular to both the current and the (out-of-plane) magnetic field. Fig.\ \ref{fig:Ni_hall} shows the results of the simultaneous LTEM and in-situ Hall experiments. The Hall voltage increases almost linearly towards saturation as expected upon applying an out-of-plane magnetic field along the magnetic hard axis of a thin film sample with strong in-plane shape anisotropy. It reaches saturation at a field amplitude of approximately $\mu_0 H_{sat} = \SI{130}{mT}$.  LTEM images acquired simultaneously at a defocus of \SI{-200}{\micro m} and different field values (relative to the saturation field $\mu_0 H_{sat}$) are also shown. The images were obtained from a region marked with a yellow rectangle in the insert overview image, and red arrows point to the corresponding field values in the Hall voltage plot below. Additional magnified sections (framed in red) are displayed below each of the LTEM images for later comparison with micromagnetic simulations. While the overall LTEM contrast is low due to the limited thickness of the sample, we observe a speckle-type contrast that successively decreases upon approaching saturation and starts to vanish at around 60\% $\mu_0 H_{sat}$.

In order to understand and validate both the observed LTEM images and anomalous Hall effect and to substantiate the reliability of the in-situ measurements, we have conducted micromagnetic simulations using the GPU-accelerated finite differences software \textit{mumax} \cite{Vansteenkiste:2014, mumax}. To this end, the geometry, morphology and chemical composition of the model structure used in the simulations were adapted to the results of our experimental characterization.

\begin{figure}[tbp]
\includegraphics[width=7cm]{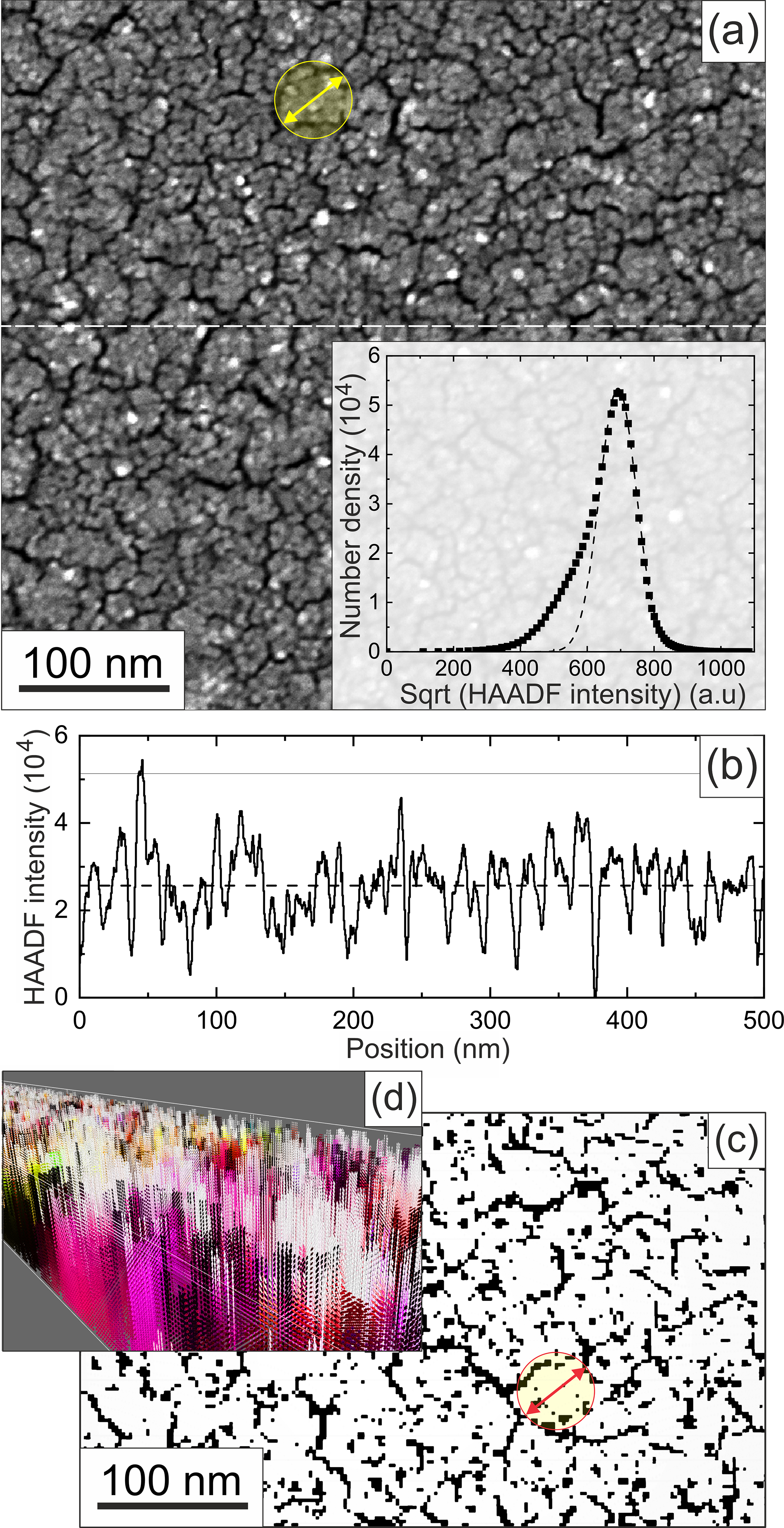}
\caption{Morphology of the Ni film: (a) STEM image of the Ni film with gray scale histogram (insert). (b) Typical line profile across the STEM image taken along the dashed line in (a). (c) Binary mask generated from the STEM image in (a). Here, white and black represent areas with and without any Ni deposit, respectively. The yellow and red arrows and circles (of identical measures) in (a) and (c) indicate the typical size of grain agglomerates. (d) Quasi-3D false-color representation of the magnetization in the film highlighting the roughness of the model structure used in the {\em mumax3} simulation.}
\label{fig:morphology}
\end{figure}

As can be seen from the bottom right insert in Fig.\ \ref{fig:Ni_hall}, the investigated Ni film had lateral dimensions of \SI{35}{\micro m} by \SI{3}{\micro m}. The RF sputter deposition of the film was adjusted to yield a nominal thickness of \SI{40}{nm}. From the STEM image in Fig.\ \ref{fig:morphology}(a) it is apparent that the film exhibits a pronounced granular morphology with individual grain and island sizes in the order of \SI{5}{nm} and \SI{50}{nm}, respectively. The mass-thickness contrast in STEM images scales roughly with the square of the atomic number Z. Therefore, the histogram in the insert of the figure, where we plot the relative frequency of the square root of the intensity measured by the high angle annular dark field (HAADF) detector, represents the height distribution of the sample. Here, a Gaussian fit (dashed line) to the peak of the data highlights that the majority of these heights are randomly distributed, however, with a slightly higher probability for smaller thicknesses. This is corroborated by the HAADF intensity profile along the dashed white line in the STEM image revealing a nanoscale roughness of the film that correlates with the individual grain size (cf.\ Fig.\ \ref{fig:morphology}(b)).

From EELS measurements, the average thickness of the Ni film was determined to be \SI{40}{nm}, which is in good agreement with the nominal thickness. Assuming a random thickness distribution and a zero minimum thickness (thereby accounting for the Ni-free paths between the grains), this translates to a thickness variation in the granular film in the range $\rm 0 \leq t \leq \SI{80}{nm}$. EELS also revealed a substantial oxidation of the granular Ni film resulting in an average composition of  $\rm Ni_{0.8}O_{0.2}$. Using the molar masses and densities of O, Ni, and NiO, and assuming that every O atom is chemically bonded to one Ni atom, this translates to a composition of \SI{25}{mole\%} NiO and \SI{75}{mole\%} Ni and results in a reduction of the volume fraction of Ni in the film to $\rm c_{vol}(Ni) = \SI{64}{\%}$. 

These findings on the film structure and chemical composition were used to set up the structure model for the micromagnetic simulations. The sample was modeled by using a core volume of $\rm \SI{3.5}{\micro m} \times \SI{0.3}{\micro m} \times thickness(nm)$ that was repeated laterally 10-fold along the x and y directions thereby representing the real sample dimensions in order to provide for realistic demagnetization fields. The granular structure of the sample was adapted from the sample by convolution of the base plane of the central volume element with a binary mask derived from a threshold STEM image (cf.\ Fig.\ \ref{fig:morphology}(c)), which was subsequently randomly subdivided into grains with planar sizes of \SI{5}{nm}. Roughness was finally introduced by adding a thickness that varies randomly from grain to grain between \SI{0}{nm} and \SI{80}{nm}. The resulting morphology of the structure model is depicted in Fig.\ \ref{fig:morphology}(d) (superimposed with a false color representation of the magnetization directions in zero field).

The room temperature saturation magnetization of Ni is $\rm M_{sat,300 K} = \SI{480}{\kA/m}$ \cite{Crangle:1971}, while NiO is antiferromagnetic and carries no net magnetization. Thus based on the oxidation-related reduction of the Ni volume, the effective saturation magnetization of the film reduces to $\rm 0.64 \times \SI{480}{kA/m} = \SI{306}{kA/m}$. Accordingly, for the micromagnetic simulations the saturation magnetization of the film was set to $\rm M_{sat} = \SI{300}{kA/m}$. The exchange stiffness is determined by renormalizing $\rm A_{ex}(0 K) = \SI{15}{pJ/m}$  \cite{Kuzmin:2020} to room temperature using $\rm A_{ex}(300 K) = A_{ex}(0 K)\times( M_S(300 K) / M_S(0 K))^2$ and set to $\rm A_{ex} = \SI{13.3}{pJ/m}$. It is established that NiO films may impose a surface anisotropy of roughly \SI{1}{mJ/m^2} \cite{Kowacz:2021}. Since surface anisotropies cannot be accounted for accurately in {\em mumax}, it was approximated by assuming a top layer with an according uniaxial volume anisotropy. For a surface layer with a thickness of one simulation cell (\SI{2.5}{nm}) and using the equality of surface and correlated effective volume anisotropy {\em energies}, $\rm E_A^{Surf}$ and $\rm E_A^{Vol}$ (i.e.\ $\rm E_A^{Surf} = \SI{3.5}{\micro m}\times\SI{0.3}{\micro m}\times\SI{1}{mJ/m^2} = \SI{3.5}{\micro m}\times\SI{0.3}{\micro m}\times\SI{2.5}{nm}\cdot K_U^{surf} = E_A^{Vol}$) this translates to an effective uniaxial anisotropy in this surface layer of $\rm K_U^{surf} = \SI{4E5}{J/m^3}$. In order to account for the fact that due to the roughness, only a minor part of the model structure reaches into this surface layer (cf.\ Fig.\ \ref{fig:morphology}(c,d)), its thickness and effective uniaxial anisotropy are slightly increased and set to \SI{7.5}{nm} (three cells) and $\rm K_U^{surf} = \SI{1E6}{J/m^3}$ in the simulations. In order to account for (i) the limited geometrical overlap of individual grains and (ii) the surface oxidation of the sample, the planar inter-grain coupling is reduced to \SI{50}{\%}  while the perpendicular exchange coupling is kept at full strength, since due to the island growth during the deposition, structural integrity along the z-axis can be assumed.

The magnetization patterns obtained from magnetic field-dependent energy minimizations using {\em mumax} were then used as input data for subsequent simulations of the LTEM images. The latter were conducted using the {\em \mbox{PyLorentz}} package \cite{McCray:2021, PyLorentz} with the appropriate parameters for the JEOL Jem-F200 (CFEG) microscope ($\rm E = \SI{200}{kV}$, $\rm C_S = \SI{95}{mm}$, $\rm C_C = \SI{18.6}{mm}$, coherence angle = \SI{0.1}{mrad}, $\rm defocus = \SI{200}{\micro m}$).

\begin{figure}[tbp]
\includegraphics[width=7cm]{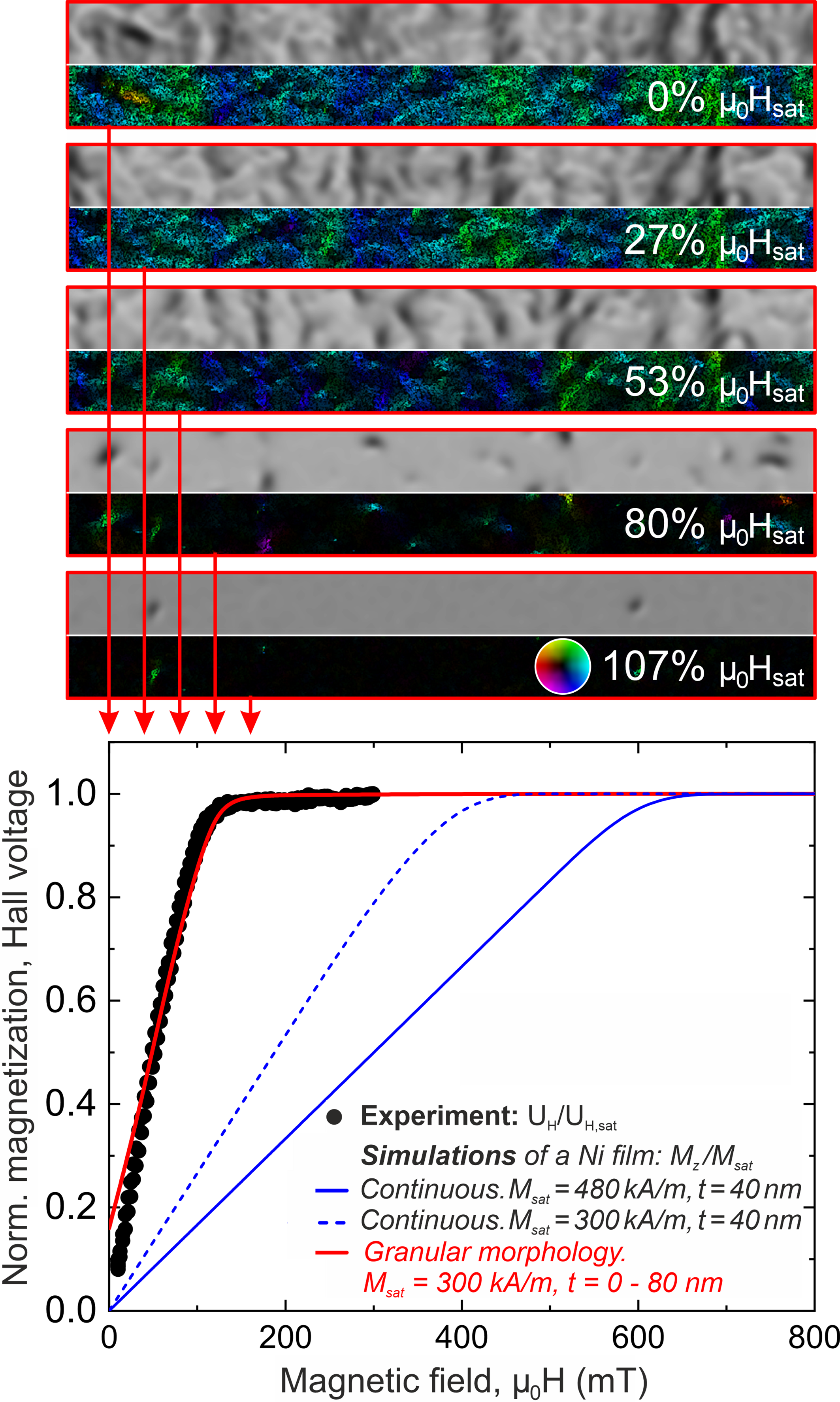}
\caption{Comparison of experiment and simulations. Bottom: In-situ measured Hall voltage, $U_{H}$ (black dots), and simulated perpendicular components of the magnetization, $M_{z}$ (solid lines), normalized to their maxima, $U_{H,sat}$ and $M_{sat}$, respectively. The pairs of images at the top represent the LTEM contrast (upper rows) and false color-coded magnetization (lower rows) for $\rm \SI{3.5}{\mu m} \times \SI{0.3}{\mu m}$ large areas as obtained from simulations based on the results of the micromagnetic {\em mumax} calculations. The in-plane components of the magnetization are directed as indicated by the color wheel in the bottom row. See text for details.}
\label{fig:simulation}
\end{figure}

The results of these simulations are summarized in Fig.\ \ref{fig:simulation}. In Ni, the AHE scales with $z$ component of the magnetization $M_z$ \cite{Kundt1893,pugh_hall_1930,pugh_hall_1932,Nagosa:2010}. Other magnetotransport phenomena (such as  anisotropic magnetoresistance), which may contribute to the signal in our geometry, were neglected because they are small compared to the AHE. Consequently, we plot at the bottom of the figure $M_z$, obtained from the {\em mumax} simulations together with the experimentally determined Hall voltage, $U_H$, both normalized to their saturation values, $M_{sat}$ and $U_{H,sat}$, respectively, as functions of the external magnetic field perpendicular to the film plane (i.e., along the z axis). Results are shown for three different scenarios: Under the assumption of a homogeneous Ni film with the nominal thickness of \SI{40}{nm} with (i) the full saturation magnetization of Ni at room temperature of $\rm M_{S} = \SI{480}{\kA/m}$ (solid blue line) and with (ii) a reduced magnetization of $\rm M_{S} = \SI{300}{\kA/m}$ that accounts for the partial oxidation (dashed blue line), as well as for (iii) a granular Ni film, whose morphology and oxidation were in the best possible way adapted to the real film structure and composition as described above (solid red line). 

It is apparent from the plot, that only by taking the real film morphology and composition as determined by STEM and EELS  into account the in-situ measured Hall voltage can be reproduced. Only this way, the most striking experimental finding -- a significant reduction of the saturation field as compared to that of a homogeneous film -- can be reproduced without any unrealistic adjustments of the simulation parameters (e.g, by a further and unjustifiably strong reduction of the saturation magnetization). Minor remaining discrepancies between experiment and simulation such as a slightly over-estimated magnetization around zero magnetic fields are due to technical limitation in setting up the sample geometry and properties using {\em mumax} that still render the model structure an {\em approximation} to the real film structure, though already a very good one.

The upper part of Fig.\ \ref{fig:simulation} shows pairs of $\rm \SI{3.5}{\mu m} \times \SI{0.3}{\mu m}$ large sections of both the simulated projected in-plane magnetization (integrated along the z-axis of the film, bottom image per pair) and the corresponding LTEM images (top image each) for external fields of 0, 40, 80, 120, and \SI{160}{mT} representing 0, 27, 53, 80, and \SI{107}{\%} of the simulated saturation field of approximately \SI{130}{mT}. The in-plane projected magnetization patterns reveal the presence of sub-micron sized domains with finite in-plane magnetization that give rise to a weak and grainy, though clearly visible domain wall contrast in the LTEM images. Upon approaching the saturation field, both the in-plane magnetic domains and the corresponding LTEM contrast vanish as the sample becomes increasingly homogeneously magnetized along the out-of-plane direction. Nonetheless, unlike the simulated images, the experimental LTEM contrast exhibits a weak remaining contrast that persists even at the saturation field. This is, however, owed to the roughness of the sample causing an inhomogeneous variation of the electrostatic potential and, thus, a remaining phase contrast that cannot be accounted for in the LTEM simulations, which are based on magnetic phase contributions only. Accordingly, the results of our simulations are in very good agreement with the experimental LTEM images. Together with the calculated $M_z(H)$ course that is consistent with the measured Hall voltage, these findings not only lend credibility to the underlying micromagnetic simulations but in turn also validate the in-situ Hall experiments in the TEM.

\subsection{Summary and conclusion}
We have successfully established a novel platform to conduct in-situ Hall effect measurements in a transmission electron microscope in combination with both field-dependent magnetic imaging using Lorentz transmission electron microscopy (LTEM) and high resolution structural and chemical characterization of the sample under investigation. The use of an external polarity switch for the current supply of the objective lens on our JEOL Jem-F200 (CFEG) microscope allows us to measure full hysteresis loops in the magnetic field range of $\SI{-1.96}{T}\leq \mu_0 H_z\leq \SI{1.96}{T}$ and in the field range of $\SI{-0.6}{T}\leq \mu_0 H_z\leq \SI{0.6}{T}$ while maintaining the automatic correction of most beam deflections. The approach is not limited to LTEM but allows also for magnetic characterization using electron holography, differential phase contrast or electron energy loss magnetic chiral dichroism (EMCD) as well as for any other (S)TEM-based technique, however at an accordingly reduced resolution in Lorentz mode.

The pre-determined changes of most beam deflections with varying OL excitation is compensated for by use of a Python script that not only controls the microscope and magneto-transport measurement hardware. It also allows to pre-define a sequence of experimental parameters thereby providing for a smooth conduction of automated measurement cycles. A commercial Prochips Fusion Select{} 
holder was used with home-made lithographically defined carrier chips for the in-situ experiments. The noise level of the Hall voltage achieved with this setup is comparable to that of dedicated magneto-resistance experiments, e.g., in a cryostat at a sample temperature of \SI{300}{K}. 

A thin Ni film has been used for proof-of-principle measurements. While Ni is in general a well understood magnet, the complex granular structure of the sputter-deposited film under investigation renders it a challenging test sample. The results of our successful in-situ experiments were confirmed by elaborated micromagnetic and subsequent LTEM image simulations. A careful comparison between experiment and simulation reveals that a consistent reproduction of the experimental results is only achieved by fully adapting the model structure in the simulations to the realistic granular and rough morphology of the partially oxidized sample. This finding highlights the large added value that goes along with such a combined magnetic, magneto-transport, and structural characterization of an {\em identical} sample thereby paving the way to a by far better understanding of the magnetic textures and their origins than any other combination of ex-situ experiments on {\em different} samples of the same material would ever allow for.

\begin{acknowledgments}
This research was funded by the Deutsche Forschungsgemeinschaft (DFG, German Research Foundation) within the framework of the priority program SPP 2137 (project ID 403503416). Support of the micromagnetic simulations through the Center for Information Services and High Performance Computing of TU Dresden is gratefully acknowledged. D.K. acknowledges the support by the Lumina fellowship LQ100102201 of the Czech Academy of Sciences and the Czech Science Foundation (grant 22-22000M).
\end{acknowledgments}

\bibliography{insitu.bib}

\end{document}